\documentclass[manuscript]{aastex}

\slugcomment{submitted to Ap. J. Letters}

\shorttitle{WMAP Age Measurement and Cosmology}
\shortauthors{L.M. Krauss}

\begin{document}


\title{Implications of the WMAP Age Measurement for Stellar Evolution and Dark Energy}


\author{Lawrence M. Krauss}
\affil{Departments of Physics and Astronomy, Case Western Reserve University, 
Cleveland OH 44106-7079}
\email{krauss@cwru.edu}



\begin{abstract}
The WMAP satellite has provided a new measurement of the age of the Universe, of $13.7 \pm 0.2$ Gyr.  A comparison of this limit with constraints from stellar evolution imply that the oldest globular clusters in the Milky Way galaxy have a reasonable probability of have formed significantly after reionization.  At the same time, one can derive a direct { \it upper limit} on the time after the big bang before globular clusters in our galaxies formed of $\approx 3$ Gyr, which significantly reduces our uncertainty since before the CMB age estimate.  The WMAP age constraint can also be shown to provide a stringent { \it lower bound} on the equation of state of dark energy.  A precise value of this lower bound would require a global analysis of the WMAP parameter constraints.  However, making conservative assumptions about allowed parameter ranges and correlations one derives a lower bound of $ w > -1.22$.  Combining this with the WMAP-quoted upper limit on $w$ thus gives roughly symmetric $95\%$ confidence range $w =-1 \pm 0.22$.

\end{abstract}




\section{Introduction}
 
The recent results reported from the WMAP CMB probe have provided a wealth of new precision data for cosmology and astrophysics.   While some of the results have been anticipated by earlier ground-based anisotropy measurements (\cite{boomerang}), several new and surprising claims have been made on the basis of the first year WMAP data (\cite{spergel}).  Particular attention has been paid to the facts that the power spectrum at low l (large angle) seems to differ somewhat from that predicted by a single power-law inflationary spectrum, and the fact that the epoch of reionization, presumably associated with the first period of star formation is at a surprisingly high redshift, corresponding to an age of approximately 200 Myr.     

At the same time, the WMAP collaboration reported a tight measurement of the age of the Universe, of $13.7 \pm 0.2$ Gyr (\cite{spergel}), by combining their data with earlier CMB and large scale structure data, supporting earlier CMB-derived age estimates (\cite{knox,boomer}) (Note that all parameters we quote here are extracted from this "best-fit" analysis from WMAP). 

While the former two observations may force revisions in our thinking about the early Universe, the latter measurement, combined with constraints on the age of globular clusters can provide new information on the formation of large scale structure, star formation, and the formation history of the Milky Way.  In addition, one can put a new independent lower bound on the equation of state parameter, $w=(p/{\rho})$, for the dark energy that appears to dominate the Universe. (\cite{krauss}).

\section{The WMAP age and the formation history of the Milky Way}

Conventional wisdom, supported by estimates of relative ages of halo vs disk clusters, suggest that halo globular clusters formed during the earliest stages of the formation of our galaxy, before the primordial gas cloud dissipated energy and collapsed to form a disk.   Thus, determination of the age of the oldest globular clusters in the halo lead to a robust lower limit on the age of the Universe (i.e. see \cite{krausschab}).     

While globular cluster ages have thus presented a good lower bound on cosmic ages, they are less successful at providing an upper limit.   This is simply because there has been no easy way to directly determine what the maximum period between the Big Bang and the formation of our own galaxy actually is.   Measurements of cosmic structure formation have suggested that galaxy formation began in earnest at redshifts less than about 7, but a minimum redshift at which it is highly likely that galaxies such as our will have formed is far less certain because no direct measurement of such a redshift has been possible.  Estimates in the range of $z \approx 1-2$ are not unreasonable, and in a cosmological constant-dominated universe this could correspond to a cosmic age of 4-5 Gyr.

By comparing WMAP observations with previous estimates of globular cluster ages, one can derive provide important new handles to probe the likely formation of the milky way galaxy, and in a broader sense the formation of large scale cosmic structures.  The two key WMAP observations in this regard are the estimate of cosmic age ($ 13.7 \pm 0.2$ Gyr), and the redshift of reionization, at $z \approx 17$ (\cite{spergel}).  

A recent comprehensive Monte Carlo analysis of the age of the oldest globular clusters that attempts to incorporate existing systematic uncertainties  yields a 68$\%$ lower confidence limit age of 11.2 Gyr (\cite{krausschab}).   Comparing this with the 68$\%$ upper limit on the age of the Universe from WMAP of $13.9$ Gyr suggests an $90\%$ upper limit  $\approx 2.7$ Gyr as the time after the Big Bang that globular clusters in our galaxy first formed from the primordial halo of gas that ultimately collapsed to form the Milky Way.  At the $ 95\%$ confidence level the limit becomes approximately 3 Gyr.   This not only improves upon previous estimates, it is the first direct constraint on this quantity.

Of somewhat more interest is a determination of the most probable time after the Big Bang at which our globular clusters formed.  Now that WMAP has determined a surprisingly early, if somewhat broad range of redshifts  near $z=17$ where the Universe reionized, corresponding to an age of about 200 Myr after the Big Bang, it is interesting to know whether this corresponds to an early period of star formation, and whether structures as large as globular clusters of stars also formed this early.  Note that Jimenez et al (\cite{jimenez}) have recently assumed this to be the case. 

A variety of different methods have been used to determine the age of globular clusters in our galaxy.  The Monte Carlo analysis referred to above involves dating these clusters using main-sequence turnoff luminosity, and yields an age estimate for the oldest clusters at the 95$\%$ confidence level of $12.6\pm ^{3.4}_{2.2}$ Gyr.  The most likely age for these globular clusters is thus $\approx$ 800 Myr younger than the WMAP lower limit on the age of the Universe.   However, because the distribution is broad, the possibility that globular clusters formed before the period of cosmic reionization determined by WMAP to occur $\approx$ 200 Myr after the Big Bang is certainly still viable (\cite{jimenez}).    Nevertheless, examining the probability distribution in \cite{krausschab}, as fit analytically in \cite{jimenez}, one finds a 75$\%$ likelihood that the oldest globular clusters are in fact less than 13.5 Gyr old.

While not compelling, the possibility that globular clusters in our galaxy may have formed well after reionization could shed light on a number of issues, including whether reionization is due to a very early generation of massive stars and whether such systems formed  before (and if so, how much before) larger structures such as globular clusters.  This could probe the nature of possible hierarchical clustering.   The likelihood of this possibility is increased when one recognizes that several other methods for determining the age of globular clusters, including using luminosity functions (\cite{jimenez2}), white dwarf cooling (\cite{hansen}) and eclipsing binaries (\cite{krausschab2}) favor globular cluster ages in the range of 11-13 Gyr.   

The existing uncertainty in globular cluster dating techniques is at present too large to do more than hint that there may be a gap in time between  reionization in the Universe and the formation of larger scale structures.  However, this hint  strongly motivates efforts to further reduce the absolute uncertainty in globular dating techniques.   In particular, the possible use of the age-mass relation suggested by Paczynski (\cite{pacz}) by observing eclipsing binaries in a number of different clusters holds great promise for reducing the absolute age uncertainty down to well below 1 Gyr, which would be required in order to firmly resolve this question.  Ultimately,  direct parallax distance measures to globular clusters will allow main sequence turnoff dating uncertainties to also fall below this level.

\section{A lower bound on $w$ }

Determination of the distance-redshift relation made using distant Type 1a supernovae (\cite{perl,kirsh}), combined with independent estimates for both the mass density
in the Universe today, and the geometry of the universe from CMB measurements (\cite{boomerang,maxima}) have definitively established the need for a dominant component to the energy budget that involves a negative pressure.  

An obvious candidate for this dark energy is a cosmological constant, with $w=-1$, but since we do not have any underlying theory for the dark energy one must allow for  the possibility that $w <-1$ (\cite{caldwell}).  Lagrangian models that have an equation of state of this form will be extremely exotic, implying for example, a negative kinetic term.  In such models energy density of the dark energy will {\it increase} with time!  As a result, the Hubble constant itself will continue to increase with time.

Age estimates can in principle give strong constraints on values of $w$ less than -1, since the age of the Universe is a strongly varying function of $w$ for values of $w >-5$ (\cite{krauss}).
In the approximation of constant $w$, which is a good approximation if $w$ is close to -1, the age relation for a flat universe is given by: 
\begin{eqnarray}
 { \nonumber H_0t_0 =  \hspace{60mm} } \\ \nonumber \\
\nonumber \int _{0}^{\infty }{dz \over{(1+z)} 
[(\Omega_{m})(1+z)^3  +  (\Omega_X)(1+z)^{3(1+w)}]^{1/2}}
\end{eqnarray}
\noindent{where $\Omega_{m}$ is the fraction of the closure density in matter today, and $\Omega_X$ is the fraction of the closure density in material with an equation of state parameter $w$.}

We display in Figs. 1 and 2, the predicted age of the Universe for various values of $w < -1$ as a function of the Hubble constant in comparison to the $2\sigma$ upper limit on the cosmic age from WMAP, for two different values of the assumed matter density today(corresponding to midpoint of  the WMAP allowed range for matter density, and the $2\sigma$ upper limit).   As is clear from these figures, for a flat universe the inferred bound on $w$ from the WMAP cosmic age limit depends sensitively on the assumed total matter density today.  

It is important to realize however that one is not free to independently vary $\Omega_m$ and $H$ in deriving bounds using the WMAP data.   These two quantities are themselves highly anti-correlated in the WMAP fit (\cite{spergel}).   As can be seen from the WMAP fits, as $w$ is decreased (for $w>-1$), the allowed range of $\Omega_m$ decreases linearly, while the allowed range of Hubble constant increases linearly.  If we assume this behavior extrapolates to values of $w <-1$, then we can use this relation to derive a conservative lower bound on $w$.   The most conservative bound on $w$ comes from assuming the largest allowed value of $\Omega_m$ for any value of H.   We fit this value using the anti-correlation described above, and fitting to the WMAP plots to derive $\Omega_m^{max} h^2 =0.309-0.243h$ within the allowed range of H .   When we include this relation explicitly for $\Omega_m$ in the cosmic age relation , we derive limits on the age of the Universe shown in Fig. 3. 

We use Fig. 3 to derive a bound on $w$.  To do this we also note that that lower bound on H derived from WMAP is correlated with the inferred value of $w$.   If we extrapolate the allowed range of H to values of $w < -1$, we find a lower bound on H as a function of $w$  shown by the thick solid line in this figure.   If we use this lower bound on H, and compare the predicted age as a function of $w$ with the WMAP upper limit, we derive a bound $w>-1.22$.   If we were instead to allow the full HST range for H in deriving this limit, the lower bound would decrease slightly to $w>-1.27$.

We emphasize that the bound we have derived here is conservative.  A tighter bound would no doubt be possible if a truly global analysis of all the WMAP data were carried out allowing $w <-1$, including possible correlations between the inferred WMAP age and other cosmic parameters.   We hope the WMAP team will perform such an analysis.  

For the moment, however, if one combines the result here with the WMAP-derived upper bound on $w$, one thus finds an allowed region $-1.22 <w< -.78$.  Since the value $w=-1$ is apparently favored within the limited range considered by the WMAP team, it seems reasonable to combine our result with the WMAP result to quote a best-fit range $ w=-1 \pm 0.22$.   It is interesting, but perhaps not surprising that the uncertainty is symmetric about the best fit value.

It thus appears that $w$ is quickly becoming constrained to lie very close to the value it would have if the dark energy is provided by a fundamental cosmological constant, or else some scalar field whose energy density is stuck in a metastable state.  Thus, as for the case of a comparison between globular cluster ages and the CMB age described earlier, progress will require reducing existing uncertainties if we are to distinguish between interesting cosmological alternatives.  The challenge for future observations that are sensitive to the value of $w$ will be to reduce the  uncertainty significantly if we are ever to be able to distinguish between a possible cosmological constant from some other exotic forms of dark energy.

The author acknowledges support from the DOE and useful conversations with John Ruhl and David Spergel.  





\begin{figure}
\plotone{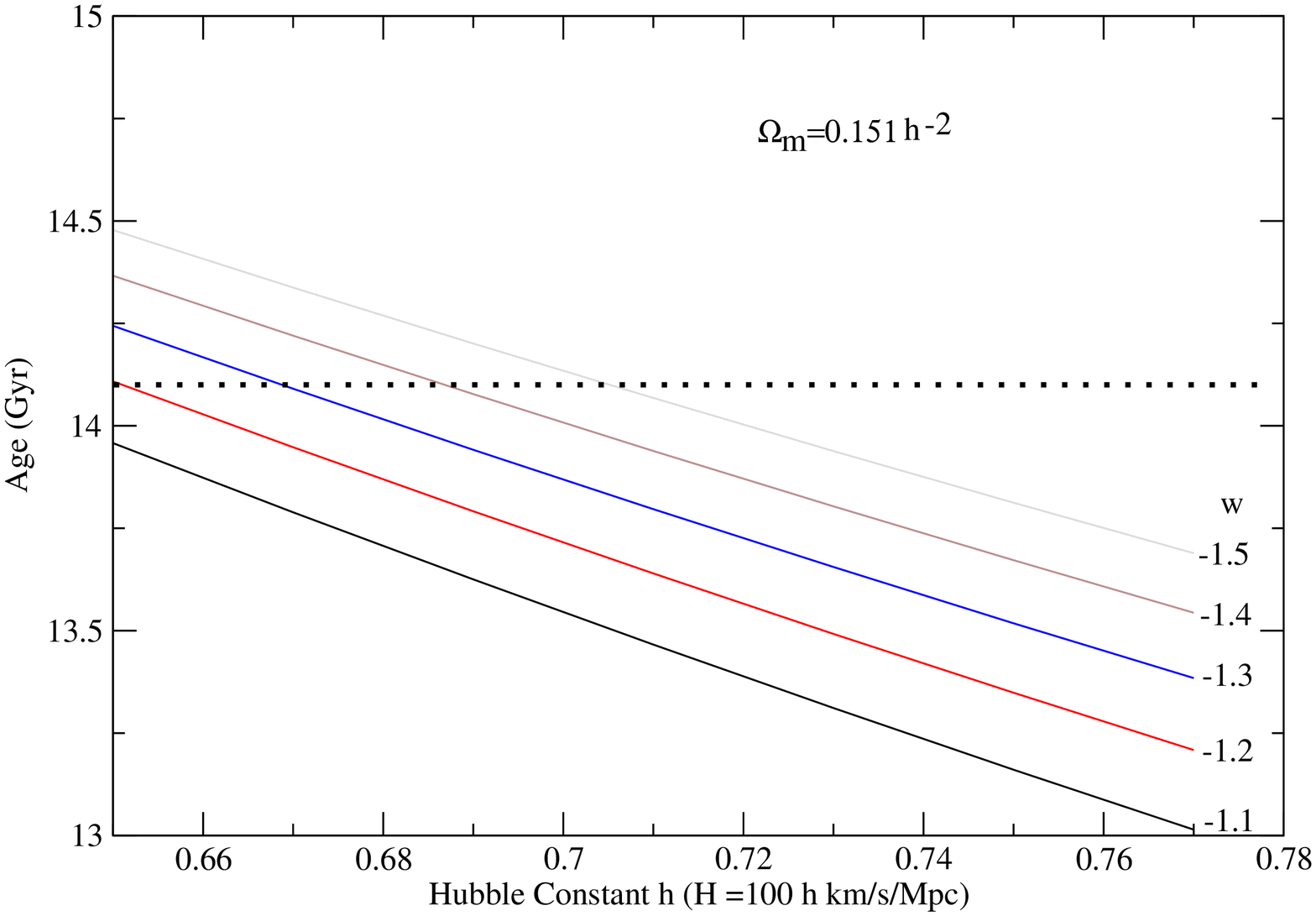}
\caption{Shown are contours of cosmic age versus Hubble constant for various constant values of the equation of state parameter for dark energy, $w <-1$, and for the matter density taking its maximum value within the $2 \sigma$ range given by WMAP, i.e. $\Omega_{m} h^2 =0.151$.  Also shown (dotted line) is the WMAP upper cosmic age constraint, where it is assumed that the age limit is not correlated with the values of  the Hubble constant within the range allowed by WMAP.  \label{fig1}}
\end{figure}

\begin{figure}
\plotone{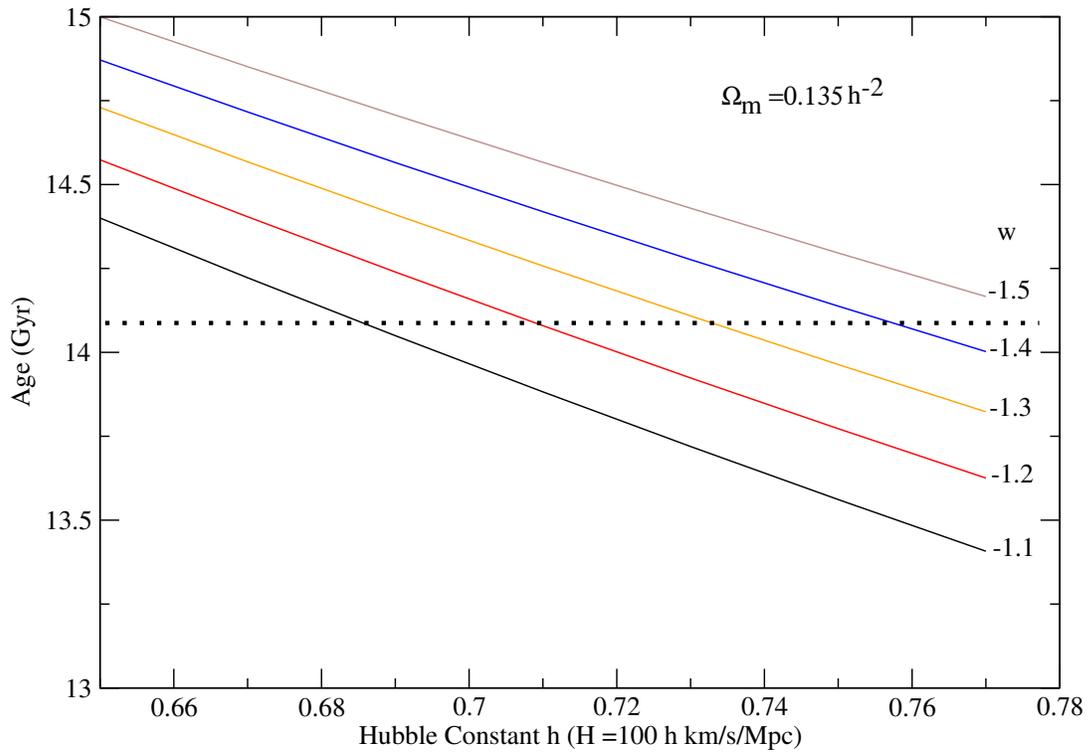}
\caption{Same as Figure 1, for the midpoint value  $\Omega_{m} h^2 =0.135$ within the $2 \sigma$ range given by WMAP.   
\label{fig2}}
\end{figure}

\begin{figure}
\plotone{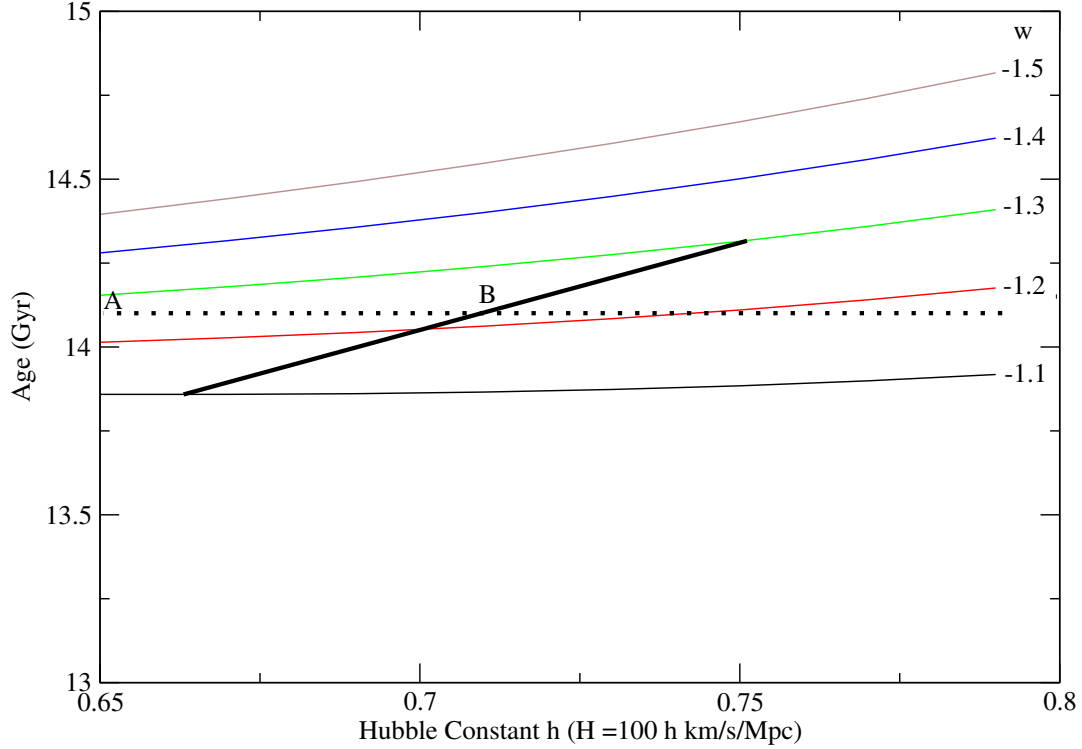}
\caption{Assuming a negative linear correlation between inferred value of the density parameter 
$\Omega_{m} h^2$ and the inferred value of $h$, based on the WMAP data, age estimates as a function of equation of state parameter $w$ can be determined.  Because predicted age is a decreasing function of the density parameter, the most conservative limits on $w$ come from choosing the maximum allowed density parameter (at the $2\sigma$ level) from WMAP. A fit to WMAP yields ($\Omega_{m}^{max} h^2=0.309-.243h$). Also shown (dotted line) is the WMAP upper cosmic age constraint, where it is assumed that the age limit is not correlated with the values of  the Hubble constant within the range allowed by WMAP.   Finally, also shown (solid curve) is the inferred lower bound on $h$ as a function of $w$ estimated by extrapolating WMAP plots.   The constraints on $w$ derived with and without this extra constraint are obtained at the points B, and A respectively.   \label{fig3}}
\end{figure}



\begin{thebibliography}{}
\bibitem[Caldwell (2002)]{caldwell} Caldwell, R., 2002, Phys. Lett. B545, 23
\bibitem[Chaboyer and Krauss (2002)]{krausschab2} Chaboyer, B., and Krauss, L.M., 2002, \apj, 567, L45
\bibitem[de Bernardis et al (2000)]{boomerang}  de Bernardis, P. {\it et al}, 2000, Nature, 404, 995
(2000)
\bibitem[Hanany et al (2002)] {maxima}Hanany, S. {\it et al}, 2000, \apj,  545, 5L 
\bibitem[Hansen et al (2002)]{hansen} Hansen, B.M. S {\it et al}, 2002, \apjl, 574, L155.
\bibitem[Goldstein et al (2002)]{boomer} Goldstein, J., et al, 2002, astro-ph/0212517
\bibitem[Jimenez and Padoan (1998)]{jimenez2} Jimenez, R., and Padoan, P., 1998, \apj 480, in press.
\bibitem[Jimenez et al (2003)]{jimenez} Jimenez, R., Verde, L., Treu, T., Stern, D., 2003, astro-ph/0302560
\bibitem[Knox et al (2001)]{knox} Knox, A., {\it et al}, 2001,  \apj, 563, L95
\bibitem[Krauss and Chaboyer (2003)] {krausschab} Krauss, L.M., and Chaboyer, B,  2003, Science, 299, 65
\bibitem[Krauss (2003)] {krauss} Krauss, L.M., 2003, astro-ph/0212369, \apj, submitted
\bibitem[Paczy\'{n}ski(1996)]{pacz}Paczy\'{n}ski, B.\ 1996, in Space   
Telescope Science Institute Series, The Extragalactic Distance Scale,   
ed.\ M.\ Livio (Cambridge: Cambridge Univ.\ Press), 273  
\bibitem[Perlmutter et al (1999)]{perl} Perlmutter, S., {\it et al},1999, \apj, 517, 565 
\bibitem[Schmidt  et al (1998)]{kirsh} Schmidt, B., {\it et al}, 1998, \apj, 507, 46
\bibitem[Spergel et al (2003)]{spergel}Spergel, D.N., 2003, {\it et al}, \apj, submitted. 
\end{thebibliography}
\end{document}